\documentclass[11pt,a4paper]{article} 
\pdfoutput=1

\usepackage{jcapmod}
\usepackage{booktabs}
\usepackage{url}
\usepackage[utf8]{inputenc}
\usepackage{enumitem}
\usepackage{amsmath}

\newcommand{\eq}[1]{eq. (\ref{#1})} 
\newcommand{\fig}[1]{figure \ref{#1}}
\newcommand{\sektion}[1]{section \ref{#1}}
\newcommand{\ocite}[1]{ref. \cite{#1}} 
 
\newcommand{\Fig}[1]{Figure \ref{#1}} 
\newcommand{\Sektion}[1]{Section \ref{#1}}

\def\ud{\mathrm{d}}

\definecolor{forfootnote}{RGB}{200,0,255}
\makeatletter
\renewcommand\@makefnmark{\hbox{\@textsuperscript{\normalfont\color{forfootnote}\@thefnmark}}}
\renewcommand\@makefntext[1]{%
  \parindent 1em\noindent
            \hb@xt@1.8em{%
                \hss\@textsuperscript{\normalfont\@thefnmark}}#1}
\makeatother

\title{Quartic inflation and radiative corrections with non-minimal coupling}

\author[a,b]{Nilay Bostan,}
\author[a,*]{and Vedat Nefer \c{S}eno\u{g}uz%
\note[*]{Corresponding author.}}

\affiliation[a]{Department of Physics, Mimar Sinan Fine Arts University, \\
Silah\c{s}\"{o}r Cad. No. 89, \.Istanbul 34380, Turkey}

\affiliation[b]{Department of Physics and Astronomy, University of Iowa, \\
Iowa City, Iowa 52242, U.S.A.}

\emailAdd{nilay-bostan@uiowa.edu}
\emailAdd{nefer.senoguz@msgsu.edu.tr}

\abstract{It is well known that the non-minimal coupling $\xi \phi^2 R$
between the inflaton and the Ricci scalar affects predictions of single
field inflation models. In particular, the $\lambda \phi^4$ quartic
inflation potential with $\xi\gtrsim0.005$ is one of the simplest models
that agree with the current data. After reviewing the inflationary
predictions of this potential, we analyze the effects of the radiative
corrections due to couplings of the inflaton to other scalar fields or
fermions. Using two different prescriptions discussed in the literature, we
calculate the range of these coupling parameter values for which the
spectral index $n_s$ and the tensor-to-scalar ratio $r$ are in agreement
with the data taken by the Keck Array/BICEP2 and Planck collaborations.}

\keywords{physics of the early universe, inflation}

\arxivnumber{1907.06215}

\begin{document} 

%comment out the line below before submitting to arXiv
%\begin{flushright}\today\end{flushright}

\maketitle \flushbottom

%adjust distance before and after equation
\setlength{\belowdisplayskip}{7.5pt} \setlength{\belowdisplayshortskip}{7.5pt}
\setlength{\abovedisplayskip}{7.5pt} \setlength{\abovedisplayshortskip}{7.5pt}

%%%%%%%%%%%%%%%%%%%%%%%%%%%%%%%%%%%%%%%%%%%%%%%%%%%%%%%%%%%%%%%%

\section{Introduction} \label{sec:intro}

Inflation \cite{Guth:1980zm,Linde:1981mu,Albrecht:1982wi,Linde:1983gd},
which is an accelerated expansion era thought to occur in the early
universe, both helps explaining general properties of the universe such as
its flatness and large scale homogeneity, and it leads to the primordial
density perturbations that evolve into the structures in the universe. Up
to now many inflationary models have been introduced with most of them
depending on a scalar field called the inflaton. Predictions of these
models are being tested by the cosmic microwave background radiation
temperature anisotropies and polarization observations that have become
even more sensitive in recent years \cite{Aghanim:2018eyx,Akrami:2018odb}.

The observational parameter values predicted by different potentials that
the inflaton may have were calculated in many articles, see for instance
ref. \cite{Martin:2013tda}. A vast majority of these articles assume that
the inflaton is coupled to gravitation solely through the metric. On the
other hand the action in general also contains a coupling term $\xi \phi^2
R$ between the Ricci scalar and the inflaton (this is required by the
renormalizability of the scalar field theory in curved space-time
\cite{Callan:1970ze,Freedman:1974ze,Buchbinder:1992rb}), and inflationary
predictions are significantly altered depending on the coefficient of this
term
\cite{Abbott:1981rg,Spokoiny:1984bd,Lucchin:1985ip,Futamase:1987ua,Fakir:1990eg,Salopek:1988qh,Amendola:1990nn,Faraoni:1996rf,Faraoni:2004pi}.

In this work, we first review in \sektion{sec:inf} how to calculate
the main observables, namely the spectral index $n_s$ and the
tensor-to-scalar ratio $r$, for an inflaton potential in the presence
of non-minimal coupling. Next, in \sektion{sec:quartic} we review the
$\lambda\phi^4$ quartic potential, providing analytical approximations
for $n_s$ and $r$, and showing that the model agrees with current data
for $\xi\gtrsim0.005$. We also briefly discuss how and to what extent
can the reheating stage affect the values of observables.   

\Sektion{sec:radiative} introduces two
prescriptions that can be used to calculate radiative
corrections to the inflaton potential due to inflaton couplings to bosons or
fermions. In prescription I, a conformal transformation is applied to
express the action in the Einstein frame; and the field dependent mass
terms in the one-loop Coleman-Weinberg potential are expressed in this
frame. Whereas in prescription II, the field dependent mass
terms are taken into account in the original Jordan frame.

The next two sections, \sektion{sec:p1} and \sektion{sec:p2} contain a
detailed numerical investigation of how the radiative corrections due to inflaton couplings to bosons or
fermions modify the predictions of the non-minimal quartic potential, for
each prescription.  We summarize our results in
\sektion{sec:conc}.

The effect of radiative corrections to the predictions of the non-minimal
quartic potential has been discussed mostly in the context of standard
model (SM) Higgs inflation \cite{Bezrukov:2007ep}, see for instance
refs. \cite{Bezrukov:2013fka,Rubio:2018ogq} and the references within. In this context, since
the self coupling $\lambda$ of the inflaton is known, $\xi\gg1$ is required
\cite{Bezrukov:2009db}. In this limit, the observational parameters are
given in terms of the e-fold number $N$ by $n_s\approx 1-2/N$ and $r\approx
12/N^2$ \cite{Komatsu:1999mt,Tsujikawa:2004my} as in the Starobinsky model
\cite{Starobinsky:1980te,Kehagias:2013mya}. Radiative corrections lead
to deviations from this so called Starobinsky point in the $n_s$ and $r$
plane, however the size of these deviations differ according to the
prescription used for the calculation. As discussed in refs.
\cite{Bezrukov:2008ej,Bezrukov:2009db,Bezrukov:2013fka}, the plateau type
structure of the Einstein frame potential remains intact and
the deviations in $n_s$ are rather insignificant according to prescription
I. However, according to prescription II, radiative corrections
lead to a linear term in the Einstein frame potential written in terms of a
scalar field with a canonical kinetic term. If the inflaton is dominantly
coupling to bosons the coefficient of this term is positive, and as this
coefficient is increased the inflationary predictions move towards the
linear potential predictions $n_s\approx1-3/(2N)$ and $r\approx4/N$
\cite{Martin:2013tda}.  If the inflaton is dominantly
coupling to fermions the coefficient of this term is negative, leading to a
reduction in the values of $n_s$ and $r$ \cite{Okada:2010jf}.

In this work we take the inflaton to be a SM singlet scalar field, and
take the self-coupling $\lambda$ and $\xi$ to be free parameters, with
$\xi\lesssim10^3$ as discussed in \sektion{sec:quartic}.\footnote{The
value of $\xi$ is ambiguous unless the inflationary part of the Lagrangian
is embedded in a specific theory (see e.g. refs.
\cite{Muta:1991mw,Faraoni:2004pi}).} Radiative
corrections for a SM singlet inflaton have been studied by refs.
\cite{Lerner:2009xg,Lerner:2011ge,Kahlhoefer:2015jma}. Unlike these
works, we focus on studying the effect of radiative corrections for
general values of $\xi\lesssim10^3$, including the case of
$\xi\ll1$.  A related work which includes the case of $\xi\ll1$ is
ref.  \cite{Okada:2010jf}. In this work the inflaton is assumed to
couple to fermions and prescription II is used. Ref.
\cite{Racioppi:2018zoy} consideres a potential which coincides with
the potential discussed in \sektion{sec:p2} for inflaton coupling to bosons.\footnote{See also
refs. \cite{Rinaldi:2015yoa,Okada:2015lia,Marzola:2015xbh,Marzola:2016xgb} for related work.} Here, we extend previous
works by considering both prescriptions I and II, and
inflaton coupling to bosons or fermions. For each case we calculate
the regions in the plane of coupling parameter values for which the spectral
index $n_s$ and the tensor-to-scalar ratio $r$ are in agreement with
the current data. We also display how $n_s$ and $r$ change due to
radiative corrections in these regions. 

Finally, we note that the non-minimal quartic inflation model given by
\eq{lagrangian} is a special case of the universal attractor models
discussed in ref. \cite{Kallosh:2013tua}. In the strong coupling limit
$\xi\to\infty$, the inflationary predictions of these models coincide with
those of conformal attractor models \cite{Kallosh:2013hoa}, which
correspond to the $\alpha=1$ case of the $\alpha$-attractor models
\cite{Kallosh:2013yoa}. The relation between these types of models is
elucidated in ref. \cite{Galante:2014ifa}.

The reheating phase of Higgs and $\alpha$-attractor-type inflation models
due to inflaton couplings to additional fields has been discussed in a
number of works, see e.g. refs.
\cite{Bezrukov:2008ut,GarciaBellido:2008ab,Bezrukov:2011gp,Ueno:2016dim,Drewes:2017fmn,Dimopoulos:2017tud}.
While the observational parameter values also depend on the details of the
reheating phase in general, for the special case of the non-minimal quartic
inflation model and for the range of $\xi$ values that we consider, the
average equation of state during reheating is given by $p\approx\rho/3$ as
we discuss in \sektion{sec:quartic}. The number of e-folds and the
observational parameter values are then to a good approximation independent
of the reheat temperature.  Thus, in our case the main effect of inflaton
couplings to additional fields on the observational parameters is not due
to the reheating phase but rather due to the radiative corrections to the
potential during inflation, which we focus on this work.

%%%%%%%%%%%%%%%%%%%%%%%%%%%%%%%%%%%%%%%%%%%%%%%%%%%%%%%%%%%%%%%%

\section{Inflation with non-minimal coupling} \label{sec:inf}

Consider a non-minimally coupled scalar field $\phi$ with a canonical kinetic
term and a potential $V_J(\phi)$:
\begin{equation} \label{vjphi}
\frac{\mathcal{L}_J}{\sqrt{-g}}=\frac12F(\phi)R-\frac12g^{\mu\nu}\partial_{\mu}\phi\partial_{\nu}\phi-V_J(\phi)\,,
\end{equation}
where the subscript $J$ indicates that the Lagrangian is specified in
Jordan frame, and $F(\phi)=1+\xi\phi^2$.  We
are using units where the reduced Planck scale $m_P=1/\sqrt{8\pi
G}\approx2.4\times10^{18}\text{ GeV}$ is set equal to unity, so we require
$F(\phi)\to1$ or $\phi\to0$ after inflation.

For calculating the observational parameters given \eq{vjphi}, it is convenient
to switch to the Einstein ($E$) frame by applying a Weyl rescaling
$g_{\mu\nu}=\tilde{g}_{\mu\nu}/F(\phi)$, so that the Lagrangian density
takes the form \cite{Fujii:2003pa}
\begin{equation} \label{LE}
\frac{\mathcal{L}_E}{\sqrt{-\tilde{g}}}=\frac12\tilde{R}-\frac{1}{2Z(\phi)}\tilde{g}^{\mu\nu}\partial_{\mu}\phi\partial_{\nu}\phi-V(\phi)\,,
\end{equation}
where
\begin{equation} \label{Zphi}
\frac{1}{Z(\phi)}=\frac32\frac{F'(\phi)^2}{F(\phi)^2}+\frac{1}{F(\phi)}\,,\quad
V(\phi)=\frac{V_J(\phi)}{F(\phi)^2}\,,
\end{equation}
and $F'\equiv\ud F/\ud\phi$. If we make a field redefinition
\begin{equation}\label{redefine}
\ud\sigma=\frac{\ud\phi}{\sqrt{Z(\phi)}}\,,
\end{equation}
we obtain the Lagrangian density for a
minimally coupled scalar field $\sigma$ with a canonical kinetic term. 

Once the Einstein frame potential is expressed in terms of the canonical
$\sigma$ field, the observational parameters can be calculated using the
slow-roll parameters (see ref. \cite{Lyth:2009zz} for a review and
references):
\begin{equation}\label{slowroll1}
\epsilon =\frac{1}{2}\left( \frac{V_{\sigma} }{V}\right) ^{2}\,, \quad
\eta = \frac{V_{\sigma \sigma} }{V}  \,, \quad
%\zeta ^{2} = \frac{V_{\sigma} V_{\sigma \sigma\sigma} }{V^{2}}\,,
\end{equation}
where $\sigma$'s in the subscript denote derivatives.
The spectral index $n_s$, the tensor-to-scalar ratio and $r$ 
%and the running of the spectral index $\alpha\equiv\mathrm{d} n_s/\mathrm{d} \ln k$ 
are given in the slow-roll approximation by
\begin{equation}\label{nsralpha1}
n_s = 1 - 6 \epsilon + 2 \eta \,,\quad
r = 16 \epsilon \,.\quad
%\alpha = 16 \epsilon \eta - 24 \epsilon^2 - 2 \zeta^2\,.
\end{equation}

The amplitude of the curvature perturbation $\Delta_\mathcal{R}$ is given by
\begin{equation} \label{perturb1}
\Delta_\mathcal{R}=\frac{1}{2\sqrt{3}\pi}\frac{V^{3/2}}{|V_{\sigma}|}\,,
\end{equation}
which should satisfy $\Delta_\mathcal{R}^2\approx 2.4\times10^{-9}$ from
the Planck measurement \cite{Aghanim:2018eyx} with the pivot scale chosen at
$k_* = 0.002$ Mpc$^{-1}$. The number of e-folds is given by
\begin{equation} \label{efold1}
N_*=\int^{\sigma_*}_{\sigma_e}\frac{V\rm{d}\sigma}{V_{\sigma}}\,, \end{equation}
where the subscript ``$_*$'' denotes quantities when the scale
corresponding to $k_*$ exited the horizon, and $\sigma_e$ is the inflaton
value at the end of inflation, which can be estimated by $\epsilon(\sigma_e) =
1$.

It is convenient for numerical calculations to rewrite these slow-roll expressions in terms of the
original field $\phi$, following the
approach in e.g. ref. \cite{Linde:2011nh}. Using \eq{redefine},
\eq{slowroll1} can be written as
\begin{equation}\label{slowroll2}  
\epsilon=Z\epsilon_{\phi}\,,\quad
\eta=Z\eta_{\phi}+{\rm sgn}(V')Z'\sqrt{\frac{\epsilon_{\phi}}{2}}\,,
%\zeta^2=Z\left(Z\xi^2_{\phi}+3{\rm
%sgn}(V')Z'\eta_{\phi}\sqrt{\frac{\epsilon_{\phi}}{2}}+Z''\epsilon_{\phi}\right)\,,
\end{equation}
where we defined 
\begin{equation}
\epsilon_{\phi} =\frac{1}{2}\left( \frac{V^{\prime} }{V}\right) ^{2}\,, \quad
\eta_{\phi} = \frac{V^{\prime \prime} }{V}  \,. 
%\zeta ^{2} _{\phi}= \frac{V^{\prime} V^{\prime \prime\prime} }{V^{2}}\,.
\end{equation}
Similarly, eqs. (\ref{perturb1}) and (\ref{efold1}) can be written as
\begin{eqnarray}\label{perturb2}
\Delta_\mathcal{R}&=&\frac{1}{2\sqrt{3}\pi}\frac{V^{3/2}}{\sqrt{Z}|V^{\prime}|}\,,\\
\label{efold2} N_*&=&\rm{sgn}(V')\int^{\phi_*}_{\phi_e}\frac{\ud\phi}{Z(\phi)\sqrt{2\epsilon_{\phi}}}\,.
\end{eqnarray}
 
To calculate the numerical values of $n_s$ and $r$ we also need a
numerical value of $N_*$. Assuming a standard thermal history after inflation, 
\begin{equation} \label{efolds}
N_*\approx64.7+\frac12\ln\frac{\rho_*}{m^4_P}-\frac{1}{3(1+\omega_r)}\ln\frac{\rho_e}{m^4_P}
+\left(\frac{1}{3(1+\omega_r)}-\frac14\right)\ln\frac{\rho_r}{m^4_P}\,.
\end{equation}
Here $\rho_{e}=(3/2)V(\phi_{e})$ is the energy density at the end of
inflation, $\rho_{*}\approx V(\phi_{*})$  is the energy density when the
scale corresponding to $k_*$ exited the horizon,  $\rho_r$ is the energy
density at the end of reheating and $\omega_r$ is the equation of state
parameter during reheating.\footnote{For a derivation of \eq{efolds} see
e.g. ref. \cite{Liddle:2003as}.} As discussed in \sektion{sec:quartic},
$\omega_r=1/3$ is generally a good approximation for the potentials which
we investigate. For this case 
\begin{equation} \label{highN} N_*\approx
64.7+\frac{1}{2}\ln \rho_*-\frac{1}{4}\ln\rho_e\,,  \end{equation}
independent of the reheat temperature.

%%%%%%%%%%%%%%%%%%%%%%%%%%%%%%%%%%%%%%%%%%%%%%%%%%%%%%%%%%%%%%%%

\section{Quartic potential} \label{sec:quartic}

Inflationary predictions of non-minimal quartic inflation have been studied
in detail, see e.g. refs.
\cite{Futamase:1987ua,Fakir:1990eg,Kaiser:1994vs,Tsujikawa:2004my,Bezrukov:2008dt,Okada:2010jf,Bezrukov:2013fca,Campista:2017ovq}.
Here after summarizing the results following ref.
\cite{Bezrukov:2013fca}, we comment on an analytical approximation used in
that work, and briefly discuss the effect of the reheating stage on the
inflationary predictions.

The Lagrangian of the non-minimal quartic inflation model is given by
\begin{equation} \label{lagrangian}
\frac{\mathcal{L}_J}{\sqrt{-g}}=\frac12(1+\xi\phi^2)R-\frac12g^{\mu\nu}\partial_{\mu}\phi\partial_{\nu}\phi-\frac14\lambda
\phi^4\,.
\end{equation}
In Einstein frame, the potential is 
\begin{eqnarray} \label{eframe}
V(\phi)=\frac{(1/4)\lambda \phi^4}{\left(1+\xi \phi^2\right)^2}\,.
\end{eqnarray}
Using eqs. (\ref{nsralpha1}) and (\ref{slowroll2}), we obtain
\begin{eqnarray} \label{nsr}
n_s & = & 1-\frac{24}{\phi^2}\left(\frac{1+\frac{5}{3}\psi+8\xi
\psi+\frac{2}{3}\psi^2+4\xi \psi^2}{(1+(1+6\xi)\psi)^2}\right)\,, 
\nonumber \\ r 
& = & \frac{128}{\phi^2}\frac{1}{1+(1+6\xi)\psi}\,,
\end{eqnarray}
where we defined $\psi\equiv\xi\phi^2$. Using \eq{perturb2} we obtain
\begin{equation} \label{lambda1}
\lambda=12\pi^2 \Delta^2_R
\frac{64(1+\psi)^2\xi^3}{\psi^2(1+(1+6\xi)\psi)}\,,
\end{equation}
and using \eq{efold2} we obtain
\begin{equation} \label{e-foldd}
N= \frac{3}{4s}(\psi-\psi_{\mathrm{e}})-\frac{3}{4}\ln
\frac{1+\psi}{1+\psi_{\mathrm{e}}}\,,
\end{equation}
where $s\equiv(6\xi)/(1+6\xi)$.
Here $\psi_{\mathrm{e}}$ can be obtained from $\epsilon(\psi_{\mathrm{e}})=
1 $ as follows:
\begin{equation} \label{end}
\psi_{\mathrm{e}}=\frac{-1+\sqrt{1+32\xi (1+6\xi)}}{2(1+6\xi)}\,.
\end{equation}
For any value of $\xi$, we can calculate the observational parameters by
numerically solving eqs. (\ref{e-foldd}) and (\ref{highN}) (with a correction for
$\xi\gtrsim1$, see below for a discussion) to find the value of $\phi_*$.
Formally, inverting \eq{e-foldd} gives a solution in terms of the $-1$
branch of the Lambert function:
\begin{equation} \label{psi_e}
\psi=-s
W_{-1}\left(-\frac{e^{-1/s}}{s\exp\left[4N/3+\psi_{\mathrm{e}}/s-\ln(1+\psi_{\mathrm{e}})\right]}\right)-1\,.
\end{equation}

\begin{figure}[tbp]
	\centering
	\includegraphics[angle=0, width=14cm]{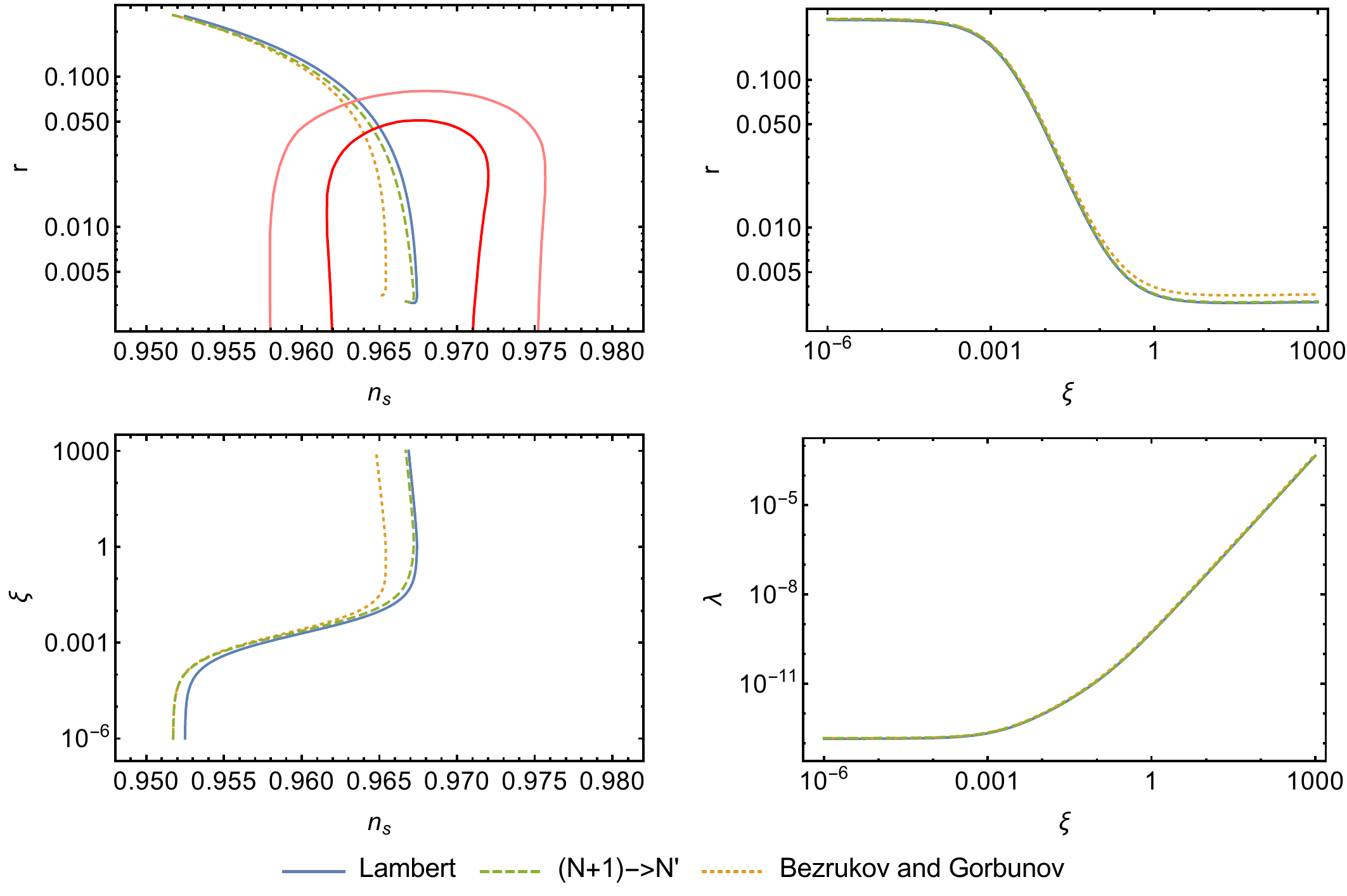}
	\caption{The values of $n_s$, $r$ and
$\lambda$ are shown as a function of the non-minimal coupling parameter
$\xi$. ``Lambert'' curves
show numerical solutions that are obtained using \eq{psi_e}, ``Bezrukov
and Gorbunov'' curves show the approximate analytical expressions in ref.
\cite{Bezrukov:2013fca}, $(N+1)\to N'$ curves show the
improved analytical expressions using \eq{n+1}. The pink (red) contour
corresponds to the 95\% (68\%) CL contour based on data taken by the Keck
Array/BICEP2 and Planck collaborations \cite{Ade:2018gkx}.}
	\label{kuartik}
\end{figure}

As \ocite{Bezrukov:2013fca} point out, one can find reasonable
approximations to the numerical solution by utilizing
$N+1\approx3\psi/(4s)$. Here we point out that a slightly more complicated
but better approximation can be obtained by using $W_{-1}(-x)\approx\ln
x-\ln(-\ln x)$:
\begin{equation}\label{approx}
\psi\approx\frac{4 s N}{3}+s\ln\left(1+\frac{4sN}{3}\right)\,.
\end{equation} 
Inserting \eq{approx} in eqs. (\ref{nsr}) and (\ref{lambda1}), we obtain
the eqs. (3.19), (3.20) and (3.22) in ref. \cite{Bezrukov:2013fca} only
with the modification
\begin{equation} \label{n+1}
N+1\to N'\equiv
N\left[1+\frac{3}{4N}\ln\left(1+\frac{4sN}{3}\right)\right]\,.
\end{equation}
Comparison of the numerical solutions and the two analytical approximations
discussed here is shown in \fig{kuartik}. As can be seen from the figure, our
analytical approximation is more accurate for $\xi\gg1$, and $n_s$ 
($r$) values deviate from
the numerical solution by at most 2\% (3\%) for any $\xi$ value. 

For the minimally coupled case, quartic potential implies an equation of
state parameter $\omega_r=1/3$
after inflation \cite{Turner:1983he}, and as a result the number of e-folds
$N_*$ is given by \eq{highN}, independent of the reheat temperature. This
result, which removes the uncertainty in the observational parameter values
due to the reheat temperature, is also valid for $\xi\lesssim1$. However, for
$\xi\gtrsim1$ the reheating stage includes a phase where the Einstein frame
potential for the canonical scalar field is quadratic
\cite{Bezrukov:2008ut,GarciaBellido:2008ab}. Following refs.
\cite{Bezrukov:2008ut,Gong:2015qha}, we obtain the e-fold number in this
case as
\begin{equation} \label{efolds2}
N_*\approx64.7+\frac12\ln\frac{\rho_*}{m^4_P}-\frac{1}{3}\ln\frac{\rho_e}{m^4_P}
+\frac{1}{12}\ln\frac{V(\phi=\sqrt2/(3^{1/4}\xi))}{m^4_P}\,.
\end{equation}
For $\xi\gtrsim1$ we use this expression instead of \eq{highN} in \fig{kuartik}. The
quadratic phase slightly reduces the value of $N_*$ with respect to the
value calculated by \eq{highN}. The difference in $N_*$ between the two
expressions is approximately given by 
\begin{equation}\label{fark} \frac{1}{12}\ln\frac{
V\left(\phi=\frac{\sqrt2}{3^{1/4}\xi}\right)}{\rho_{e}}\,.
\end{equation}
Since $\xi \phi^2\lesssim1$, $V(\phi)\propto \phi^4$ after inflation. In
this case $\rho_e$ is proportional to $\phi_e^4=4/(3\xi^2)$ for $\xi\gg1$.
Thus, for $\xi\gg1$, the reduction in $N_*$ due to the quadratic phase is
approximately given by $-(1/6)\ln\xi$ \cite{Gong:2015qha}.

For the non-minimal quartic potential, expanding the
action around the vacuum reveals a cut-off scale $\Lambda= 1/\xi$
\cite{Burgess:2009ea,Barbon:2009ya,Hertzberg:2010dc}. From eqs. (\ref{perturb1}) and (\ref{nsralpha1}), we obtain $V=(3/2)\pi^2
r\Delta_\mathcal{R}^2$. Thus, requiring $\Lambda$
to be higher than the energy scale during inflation corresponds to
\begin{equation}
\xi<\left(\frac32\pi^2 r\Delta_\mathcal{R}^2\right)^{-1/4}\,,
\end{equation} 
leading to $\xi\lesssim10^{2.5}$. For this reason all numerical results in
the following sections will be displayed for $\xi$ values up to $10^3$.
With this constraint, taking into account the quadratic phase after
inflation makes only a $\lesssim1$ difference in the $N_*$ value.
Furthermore, preheating effects can make this difference even smaller. Thus,
the uncertainty in the observational parameter values due to the reheating
stage is rather small for non-minimal quartic inflation. For instance, in
\fig{kuartik} which shows $n_s$ and $r$ values for $\xi$ up to $10^3$, the
effect of the quadratic phase amounts to the barely visible hook-like part
at the bottom end of the $n_s$--$r$ curves. Therefore, we will use the
$\omega_r=1/3$ approximation hereafter.

Finally, note that since we compare the numerical $n_s$ and $r$ values
with the recent Keck Array/BICEP2 and Planck data \cite{Ade:2018gkx}, our
constraints on $\xi$ are slightly more stringent compared to earlier works.
Namely, non-minimal quartic inflation is compatible with the Keck
Array/BICEP2 and Planck data for $\xi\gtrsim0.005$ (0.01) at $95\%$
($68\%$) confidence level (CL).

%%%%%%%%%%%%%%%%%%%%%%%%%%%%%%%%%%%%%%%%%%%%%%%%%%%%%%%%%%%%%%%%

\section{Radiative corrections} \label{sec:radiative}

Interactions of the inflaton with other fields, required for efficient
reheating, lead to radiative
corrections in the inflaton potential. These corrections can be expressed
at leading order as follows \cite{Coleman:1973jx}:
\begin{eqnarray} \label{cw1loop} \Delta
V(\phi)=\sum\limits_{i}\frac{(-1)^F}{64\pi^2}M_i(\phi)^4
\ln\left(\frac{M_i(\phi)^2}{\mu^2}\right)\,,  \end{eqnarray}
where $F$ is $+1$ $(-1)$ for bosons (fermions), $\mu$ is a renormalization
scale and $M_i(\phi)$ denote field dependent masses.

First let us consider the potential terms for a minimally coupled
inflaton field with a quartic potential, which couples to another scalar
$\chi$ and to a Dirac fermion $\Psi$:
\begin{eqnarray}\label{lmin} V(\phi,\chi,\Psi)=
\frac{\lambda}{4}\phi^4+h\phi\bar{\Psi}\Psi+m_\Psi\bar{\Psi}\Psi+\frac{1}{2}g^2\phi^2\chi^2+\frac{1}{2}m_\chi^2\chi^2\,.  \end{eqnarray}
Under the assumptions
\begin{eqnarray} \label{varsay1}
g^2\phi^2\gg m_\chi^2 \,, \quad g^2\gg\lambda \,, \quad  h\phi\gg
m_\Psi \,, \quad h^2\gg\lambda \,,  \end{eqnarray} 
the inflaton potential including the Coleman-Weinberg (CW) one-loop corrections
given by \eq{cw1loop} take the form:
\begin{eqnarray}\label{radpot3}
V(\phi)=\frac{\lambda}{4}\phi^4\pm\kappa\phi^4\ln\left(\frac{\phi}{\mu}\right)\,,
\end{eqnarray}
where the $+$ ($-$) sign corresponds to the case of the inflaton
dominantly coupling to bosons (fermions) and we have defined the radiative correction coupling parameter
\begin{eqnarray}\label{kappatanim}
\kappa\equiv\frac{1}{32\pi^2}\Big|(g^4-4h^4)\Big|\,.
\end{eqnarray}

Generalizing \eq{radpot3} to the non-minimally coupled case is subject to
ambiguity unless the ultraviolet completion of the low-energy effective
field theory is specified, as discussed in refs.
\cite{Bezrukov:2008ej,Bezrukov:2009db,Bezrukov:2013fka,Hamada:2016onh}. In
the literature, typically two prescriptions for the calculation of
radiative corrections are adopted. In prescription I, the
field dependent masses in the one-loop CW potential are expressed in the
Einstein frame.  Using the transformations
\begin{equation}
V(\phi)=\frac{V_J(\phi)}{F(\phi)^2}, \;
\tilde{\phi}=\frac{\phi}{\sqrt{F(\phi)}}, \;
\tilde{\Psi}=\frac{\Psi}{F(\phi)^{3/4}}, \; 
\tilde{m}_\Psi(\phi)=\frac{m_\Psi(\phi)}{\sqrt{F(\phi)}}, \;
\tilde{m}_\chi^2=\frac{m_\chi^2}{F(\phi)}\,,  \end{equation}
the one-loop corrected potential is obtained in the Einstein frame as
\begin{equation}\label{radpotyontem1} V(\phi)=\frac{\frac{\lambda}{4
}\phi^4\pm\kappa \phi^4 \ln\left(\frac{\phi}{\mu \sqrt{1+\xi
\phi^2}}\right)}{(1+\xi\phi^2)^2}\,.  \end{equation} 
In prescription II, the field dependent
masses in the one-loop CW potential are expressed in the Jordan frame, so
that \eq{radpot3} corresponds to the one-loop corrected potential in the
Jordan frame. Therefore the Einstein frame potential in this case is given
by
\begin{eqnarray}\label{radpotyontem2}
V(\phi)=\frac{\frac{\lambda}{4 }\phi^4\pm\kappa \phi^4
\ln\left(\frac{\phi}{\mu}\right)}{(1+\xi\phi^2)^2}\,.  \end{eqnarray}

Note that the potentials in eqs. (\ref{radpotyontem1}) and
(\ref{radpotyontem2}) are approximations that can be
obtained from the one-loop renormalization group improved effective
actions, see for instance ref. \cite{Okada:2010jf} for a discussion of this
point. 

In the next two sections, we numerically investigate how the $n_s$ and $r$
values change as a function of the coupling parameters $\xi$ and $\kappa$
using prescription I and prescription II, respectively. The calculation
procedure is as follows: We form a grid of points in the $\xi$ and $\kappa$
plane. For each $(\xi,\;\kappa)$ point, we start the calculation by
assigning an initial $\lambda$ value. We then calculate numerical values of
$\phi_e$ using $\epsilon(\phi_e) =1$, and $\phi_*$ using \eq{perturb2}. The
e-fold number $N_*$ is calculated using \eq{efold2} and compared with
\eq{highN}. The initial value of $\lambda$ is then adjusted and the
calculation is repeated until the two $N_*$ values match. The $\phi_*$
value obtained this way is plugged in eqs. (\ref{slowroll2}) and
(\ref{nsralpha1}) to yield the $n_s$ and $r$ values. Finally, the
calculation is repeated over the whole grid, with $\lambda$ solutions for
each point used as initial values of their neighbors.

For the numerical calculations, a value for $\mu$ should also be specified.
However, shifting the value of $\mu$ does not change the forms of  eqs.
(\ref{radpotyontem1}) and (\ref{radpotyontem2}), corresponding only to a
shift in the value of $\lambda$. Thus, for fixed values of the coupling
parameters $\xi$ and $\kappa$, $n_s$ and $r$ values do not depend on $\mu$.

%%%%%%%%%%%%%%%%%%%%%%%%%%%%%%%%%%%%%%%%%%%%%%%%%%%%%%%%%%%%%%%%

\section{Radiatively corrected quartic potential: Prescription I}\label{sec:p1}

In this section we numerically investigate how the $n_s$ and $r$ values
change as a function of the coupling parameters $\xi$ and $\kappa$, using
the potential in eq. (\ref{radpotyontem1}), with a $+$ ($-$) sign for the
inflaton dominantly coupling to bosons (fermions).

\begin{figure}[t!]
   	\begin{center}
\includegraphics[width=7.5cm]{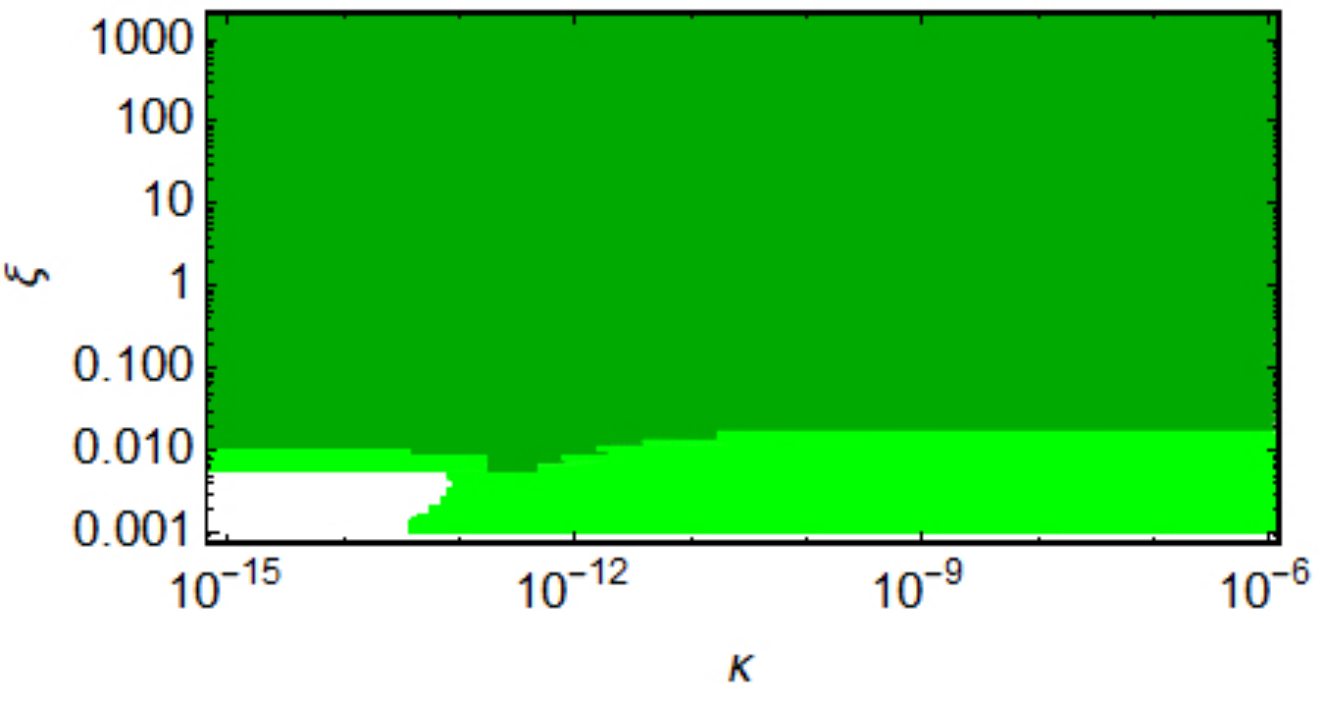}\\
\includegraphics[width=14cm]{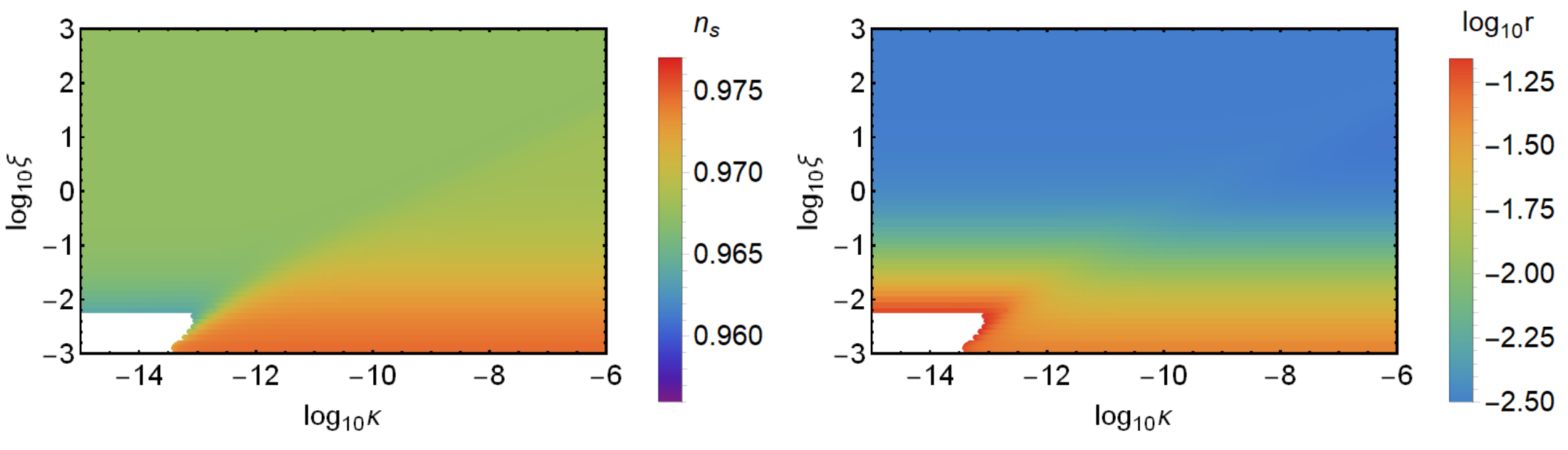}
	\end{center}\vspace*{-.5cm}
	\caption{For prescription I and inflaton coupling to bosons, the
top figure shows in light green (green) the regions in the $\xi$--$\kappa$
plane for which $n_s$ and $r$ values are within the $95\%$ $(68\%)$ CL
contours based on data taken by the Keck Array/BICEP2 and Planck
collaborations \cite{Ade:2018gkx}.  Bottom figures show $n_s$ and $r$
values in these regions.}
	\label{6.5}
\par\vspace{\intextsep}
   	\begin{center}
		\includegraphics[width=9cm]{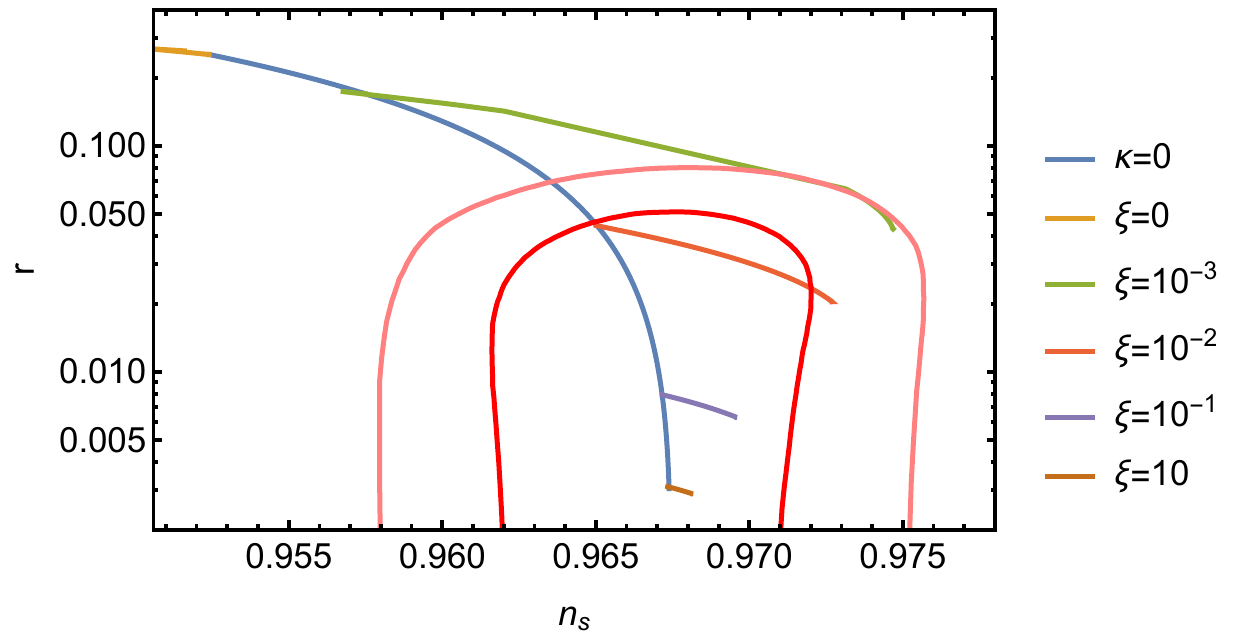}\\
		\includegraphics[width=14cm]{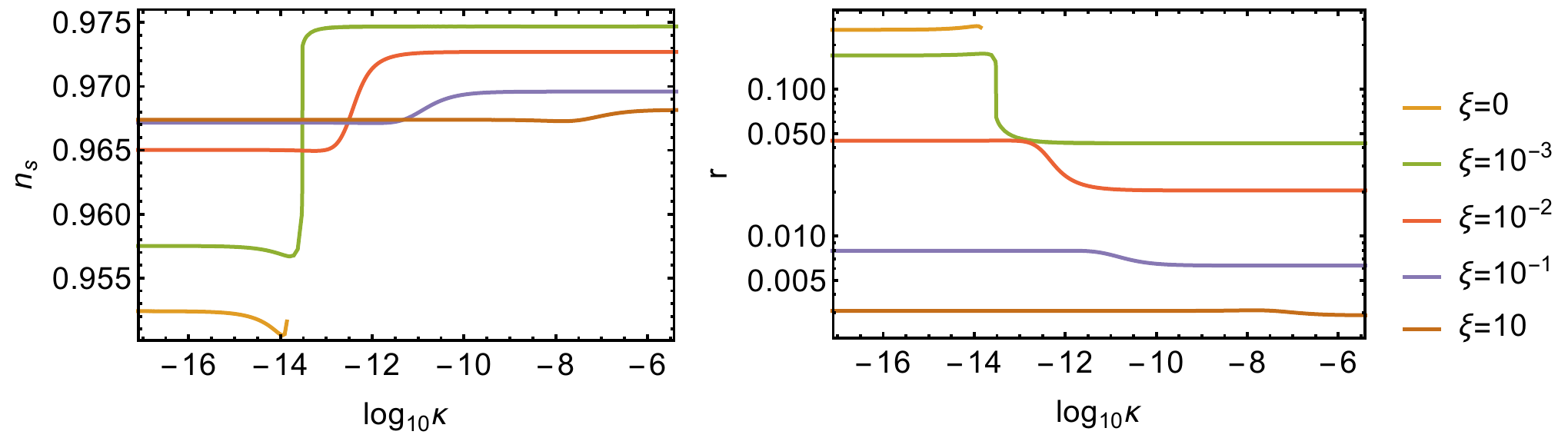}
	\end{center}\vspace*{-.5cm}
	\caption{For prescription I and inflaton coupling to bosons, the
change in $n_s$ and $r$ as a function of $\kappa$ is plotted for selected
$\xi$ values. The pink (red) contour in the top figure corresponds to the
95\% (68\%) CL contour based on data taken by the Keck Array/BICEP2 and
Planck collaborations \cite{Ade:2018gkx}.}
	\label{6.9}
\end{figure}

For prescription I and inflaton coupling to bosons,
\fig{6.5} shows the region in the $\xi$ and $\kappa$ plane where $n_s$ and
$r$ values are compatible with the current data. \Fig{6.9} shows how $n_s$
and $r$ values change with $\kappa$ for chosen $\xi$ values. It is clear
from the figures that $n_s$ and $r$ values depend more sensitively on the
value of $\xi$ rather than $\kappa$. As $\kappa$ is increased
holding $\xi$ fixed, there is a transition in $n_s$ and $r$ values for a
relatively narrow range of $\kappa$. $n_s$ and $r$ no longer change at even
larger $\kappa$ values, however this last result is subject to some caveats
as discussed below.

In contrast to the
other cases covered in subsequent sections, we find that eqs.
(\ref{perturb2}), (\ref{efold2}) and (\ref{highN}) can be simultaneously
satisfied for arbitrarily large values of $\kappa$. However, 
as mentioned in \sektion{sec:radiative}, the
potential we use is an approximation of the one-loop renormalization group
improved effective action, and this approximation will eventually fail for
large values of $\kappa$. Furthermore, higher loop corrections will eventually
become also important. 

Even if we take the potential in \eq{radpotyontem1} at
face value, the inflationary solutions for large $\kappa$ values can only
be obtained for fine tuned values of the coupling parameters. To show this,
let us write the potential in the limit $\xi\phi^2\gg1$ and take $\mu=1$
for convenience. The potential then approximately takes the form
\eq{eframe}, with $\lambda/4$ replaced by
$A\equiv\lambda/4-(\kappa/2)\ln\xi$. Using \eq{redefine}, this potential
can be written as
\begin{equation} \label{lambdakappa}
V(\sigma)\approx \frac{A}{\xi^2}\left[1-2\exp\left(-2\sqrt{\frac s6}
\sigma\right)\right]\,.  \end{equation}
Using \eq{efold1}, $\exp(2\sqrt{s/6}\sigma)\approx4sN/3$.
Finally, using \eq{perturb1}, we obtain
\begin{eqnarray} \label{lambdafine} \lambda\approx \frac{72\pi^2
\Delta^2_R \xi^2}{sN^2}+2\kappa \ln\xi\,.  \end{eqnarray}
The first term in the right hand side is approximately
$5\times10^{-10}\xi^2$ for $\xi\gg1$. If $2\kappa\ln\xi$ is much larger
than this term, \eq{lambdafine} can only be satisfied if
$\lambda$ almost exactly equals $2\kappa\ln\xi$. 

The case of inflaton having a quartic potential with radiative corrections
due to coupling to fermions was discussed in ref. \cite{NeferSenoguz:2008nn}
taking $\xi=0$. There it was pointed out that there are two solutions for
every $\kappa$ value that is smaller than a maximum
$\kappa_{\mathrm{max}}$ value. This is also true for $\xi\ne0$, with
$\kappa_{\mathrm{max}}$ values depending on $\xi$. We label the branch of
solutions with larger $\lambda$ for a given $\kappa$ as the first branch,
and the other branch of solutions as the second branch. For
$\kappa>\kappa_{\mathrm{max}}$ there is no solution, that is,  eqs.
(\ref{perturb2}), (\ref{efold2}) and (\ref{highN}) cannot be simultaneously
satisfied.

\begin{figure}[t!]
   	\begin{center}
		\includegraphics[width=7.5cm]{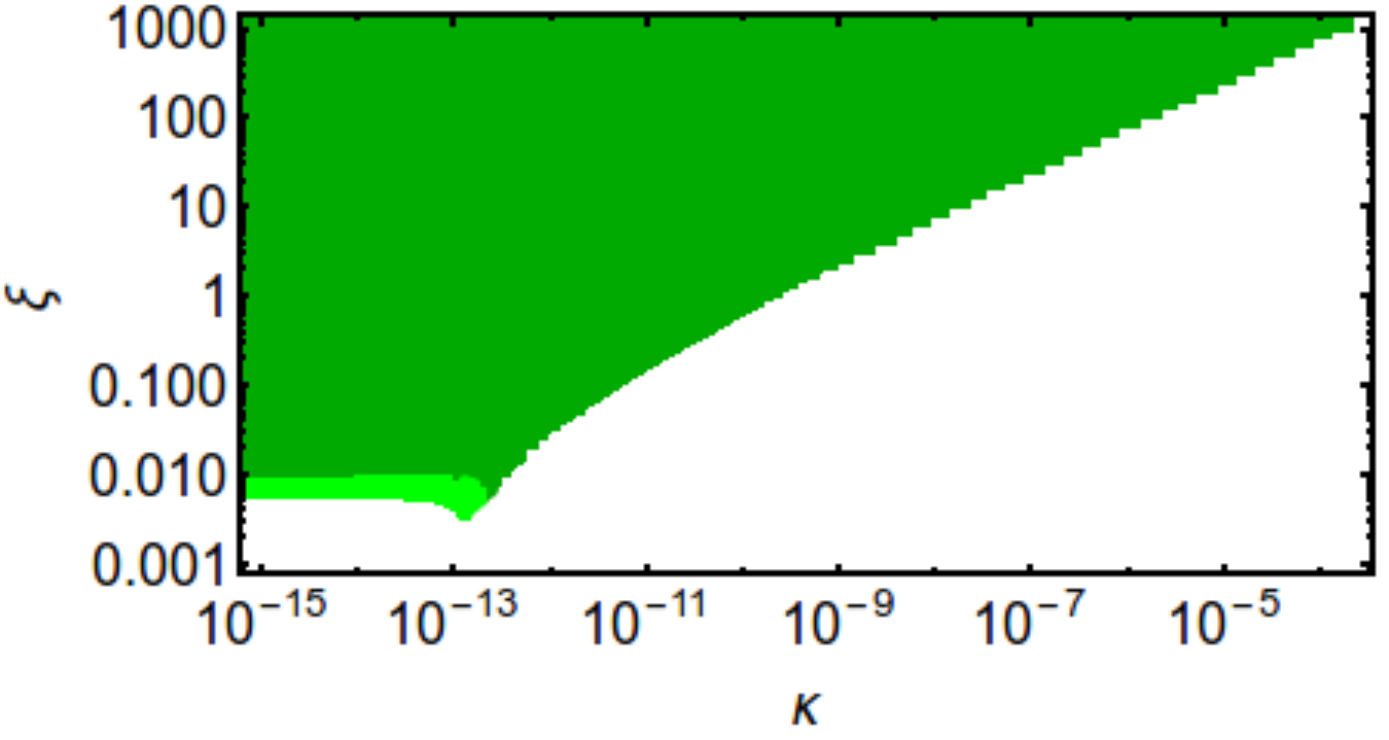}\\
		\includegraphics[width=14cm]{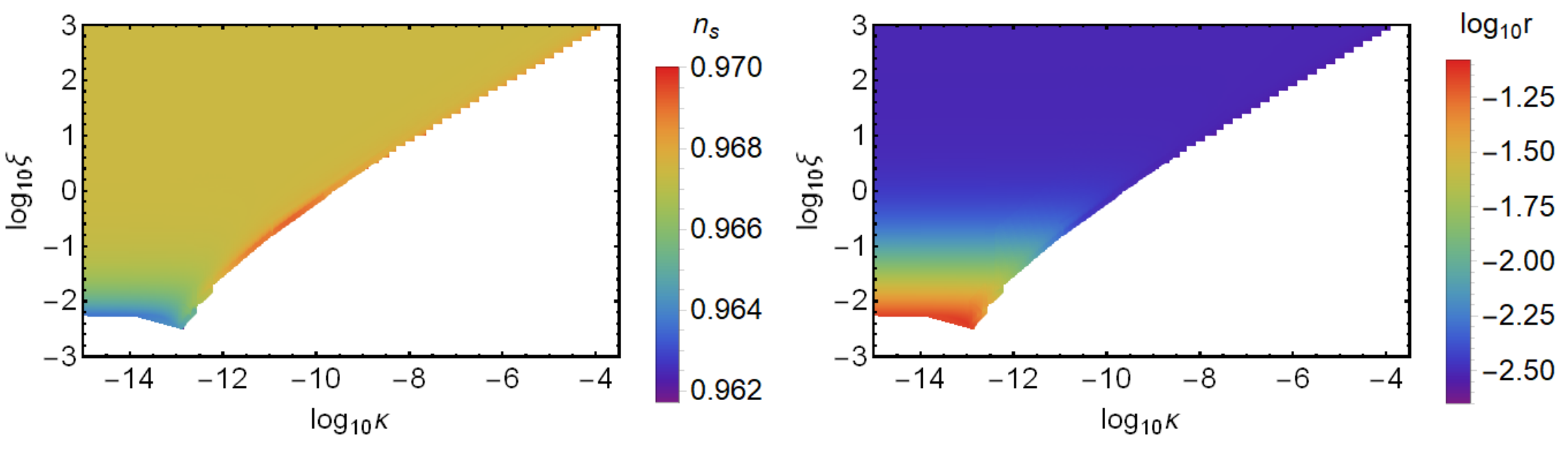}
	\end{center}\vspace*{-.5cm}
	\caption{For prescription I, inflaton coupling to fermions and
first branch solutions, the top figure shows in light green (green) the
regions in the $\xi$--$\kappa$ plane for which $n_s$ and $r$ values are
within the $95\%$ $(68\%)$ CL contours based on data taken by the Keck
Array/BICEP2 and Planck collaborations \cite{Ade:2018gkx}.  Bottom figures
show $n_s$ and $r$ values in these regions.}
	\label{6.1}
\par\vspace{\intextsep}
   	\begin{center}
		\includegraphics[width=7.5cm]{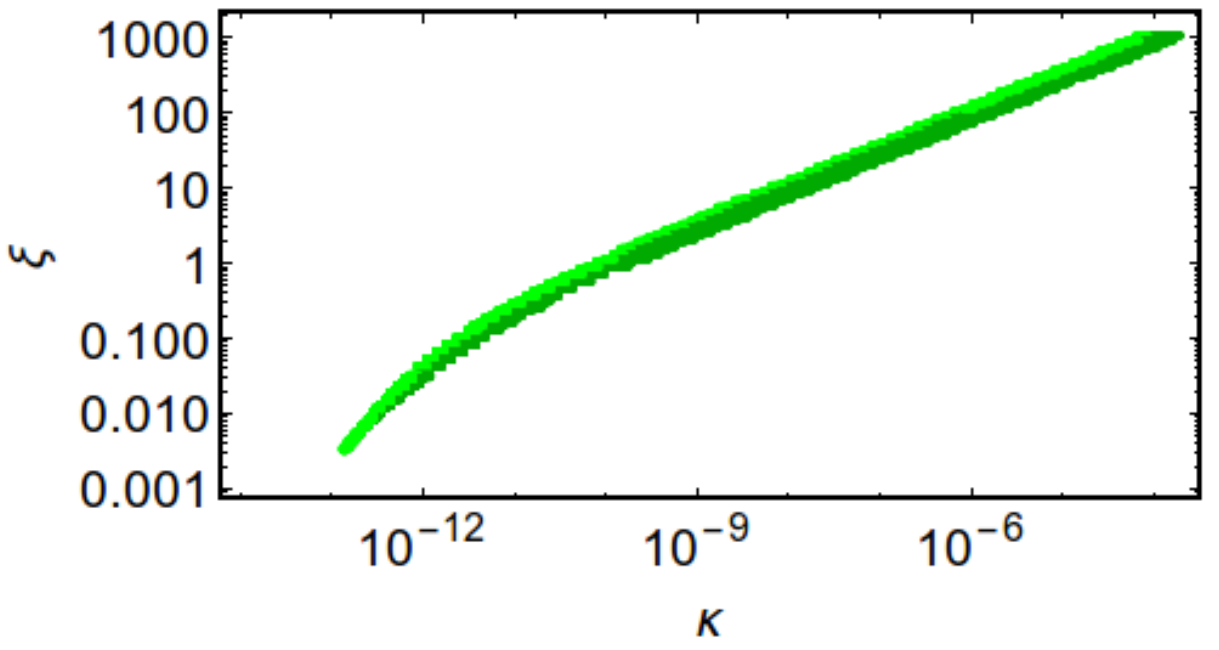}\\
		\includegraphics[width=14cm]{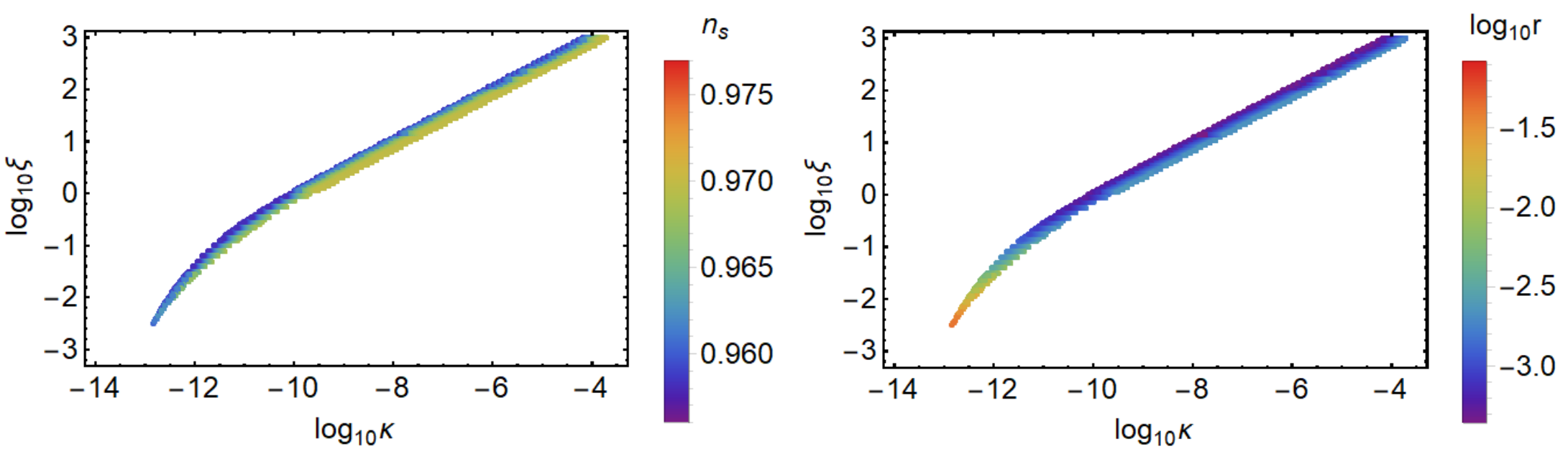}
	\end{center}\vspace*{-.5cm}
	\caption{For prescription I, inflaton coupling to fermions and
second branch solutions, the top figure shows in light green (green) the
regions in the $\xi$--$\kappa$ plane for which $n_s$ and $r$ values are
within the $95\%$ $(68\%)$ CL contours based on data taken by the Keck
Array/BICEP2 and Planck collaborations \cite{Ade:2018gkx}.  Bottom figures
show $n_s$ and $r$ values in these regions.}
	\label{6.3}\end{figure}

For prescription I and inflaton coupling to fermions,
\fig{6.1} shows the region in the $\xi$ and $\kappa$ plane where $n_s$ and
$r$ values are compatible with the current data, for the first branch of
solutions. Again, $n_s$ and $r$ values depend more sensitively on the
value of $\xi$ rather than $\kappa$. The observationally compatible region for the second
branch of solutions is shown in \fig{6.3}. As seen from the figure, the
second branch solutions are compatible with observations for only a narrow
region in the $\xi$--$\kappa$ plane.

\begin{figure}[t]
	\begin{center}
		\includegraphics[width=10cm]{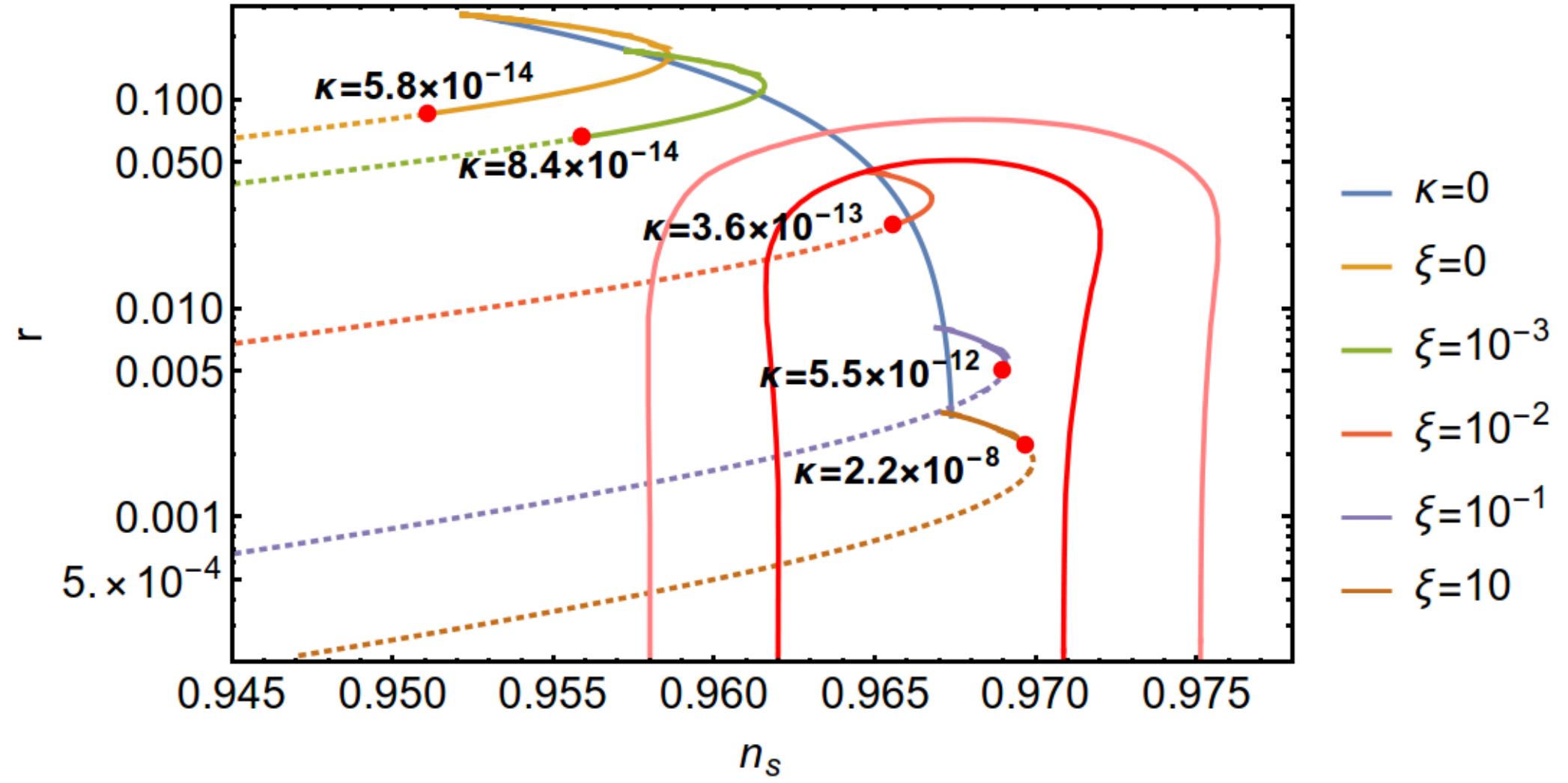}
		\includegraphics[width=14cm]{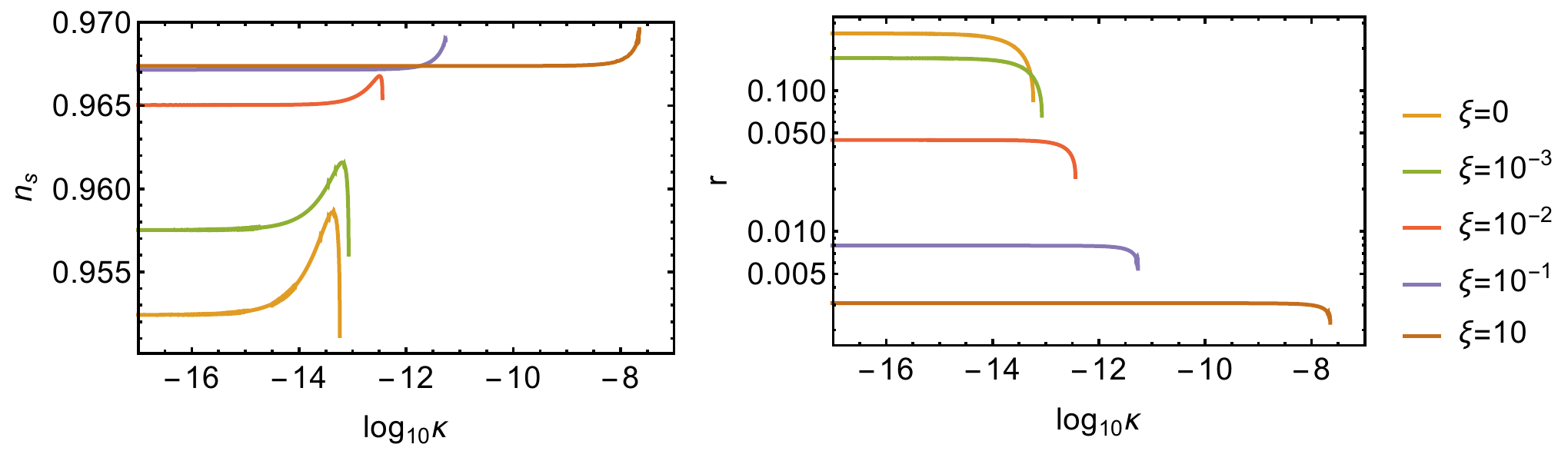}
	\end{center}\vspace*{-.5cm}
	\caption{For prescription I and inflaton coupling to fermions, the
change in $n_s$ and $r$ as a function of $\kappa$ is plotted for selected
$\xi$ values. The pink (red) contour in the top figure corresponds to the
95\% (68\%) CL contour based on data taken by the Keck Array/BICEP2 and
Planck collaborations \cite{Ade:2018gkx}. The solid (dotted) portions of
the curves correspond to first (second) branch of solutions. The red points
show the maximum $\kappa$ values where the two branch of solutions meet.
These values are also written in the figure. The
bottom figures only show the first branch solutions.}
	\label{6.11}
\end{figure}

\Fig{6.11} shows how $n_s$ and $r$ values change with $\kappa$ and the
$\kappa_{\mathrm{max}}$ values for chosen $\xi$ values. The first branch
solutions move from the red points towards the $\kappa=0$ curve as $\kappa$
decreases. As can be seen from the bottom panels, significant change in the
$n_s$ and $r$ values only occur when $\kappa$ becomes the same order of
magnitude as $\kappa_{\mathrm{max}}$. The second branch solutions, on the
other hand, move towards small $n_s$ values and away from the
observationally favored region in the $n_s$--$r$ plane as $\kappa$
decreases. These solutions cease to exist for
$\kappa\ll\kappa_{\mathrm{max}}$ as inflation with sufficient duration
cannot be obtained.

%%%%%%%%%%%%%%%%%%%%%%%%%%%%%%%%%%%%%%%%%%%%%%%%%%%%%%%%%%%%%%%%

\section{Radiatively corrected quartic potential: Prescription II}\label{sec:p2}

In this section we numerically investigate how the $n_s$ and $r$ values
change as a function of the coupling parameters $\xi$ and $\kappa$, using
the potential in eq. (\ref{radpotyontem2}), with a $+$ ($-$) sign for the
inflaton dominantly coupling to bosons (fermions).

%\enlargethispage{\baselineskip}

\begin{figure}[t]
	\begin{center}
		\includegraphics[width=7.5cm]{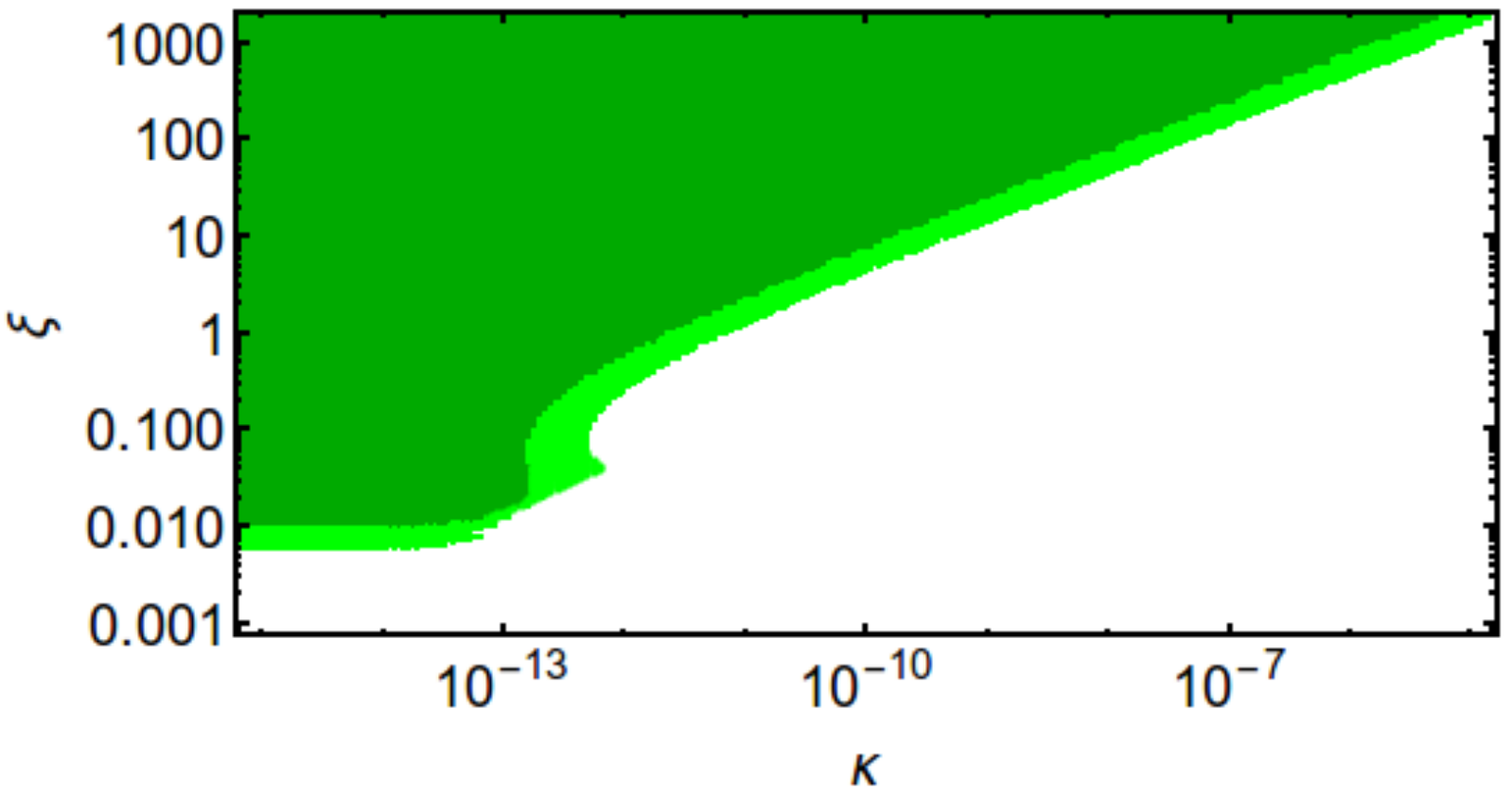}\\
		\includegraphics[width=14cm]{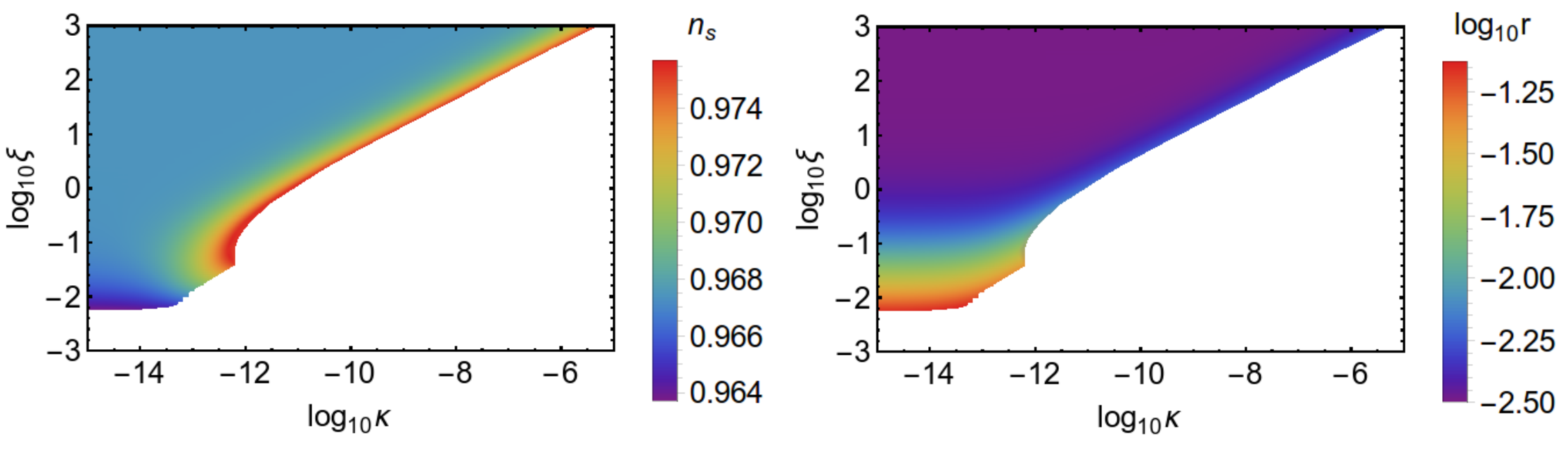}
	\end{center}\vspace*{-.5cm}
	\caption{For prescription II and inflaton coupling to bosons, the top figure shows in light green (green) the
regions in the $\xi$--$\kappa$ plane for which $n_s$ and $r$ values are
within the $95\%$ $(68\%)$ CL contours based on data taken by the Keck
Array/BICEP2 and Planck collaborations \cite{Ade:2018gkx}.  Bottom figures
show $n_s$ and $r$ values in these regions.}
	\label{7.3}
\end{figure}

\begin{figure}[b]
	\begin{center}
		\includegraphics[width=9cm]{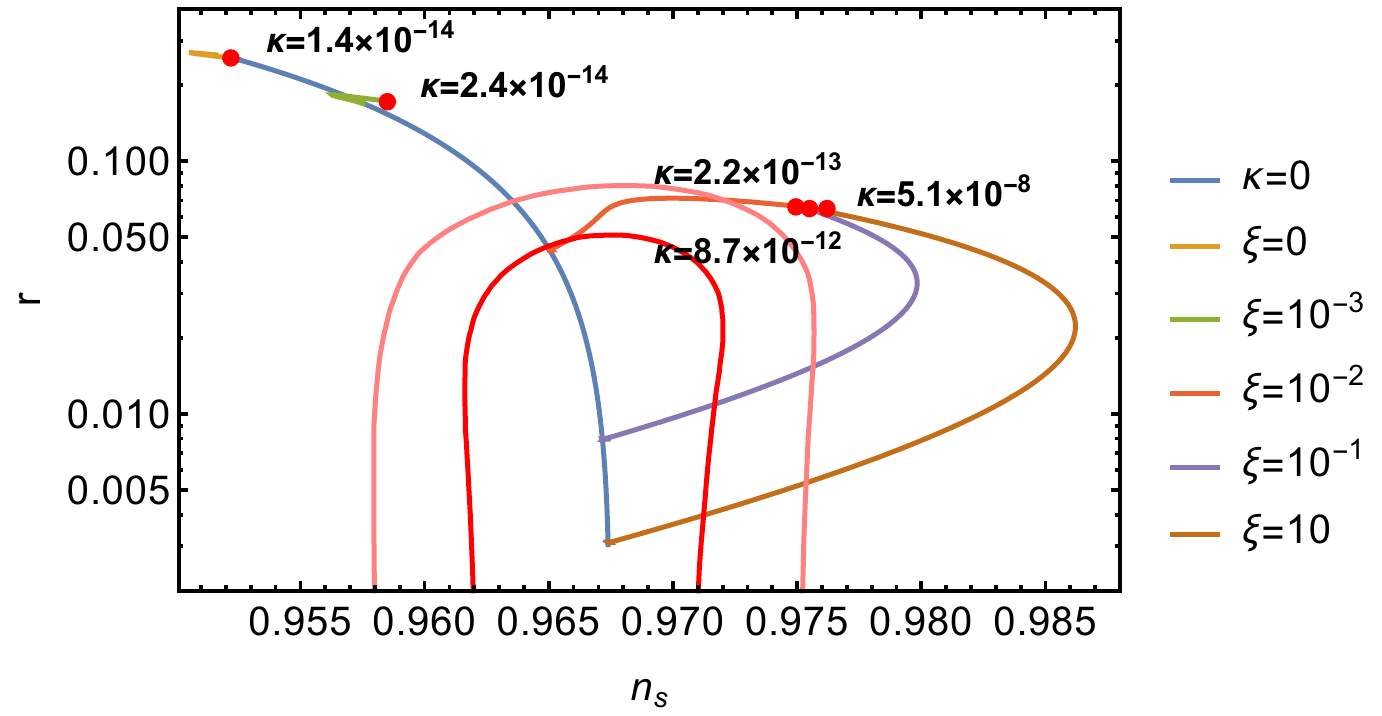}\\
		\includegraphics[width=14cm]{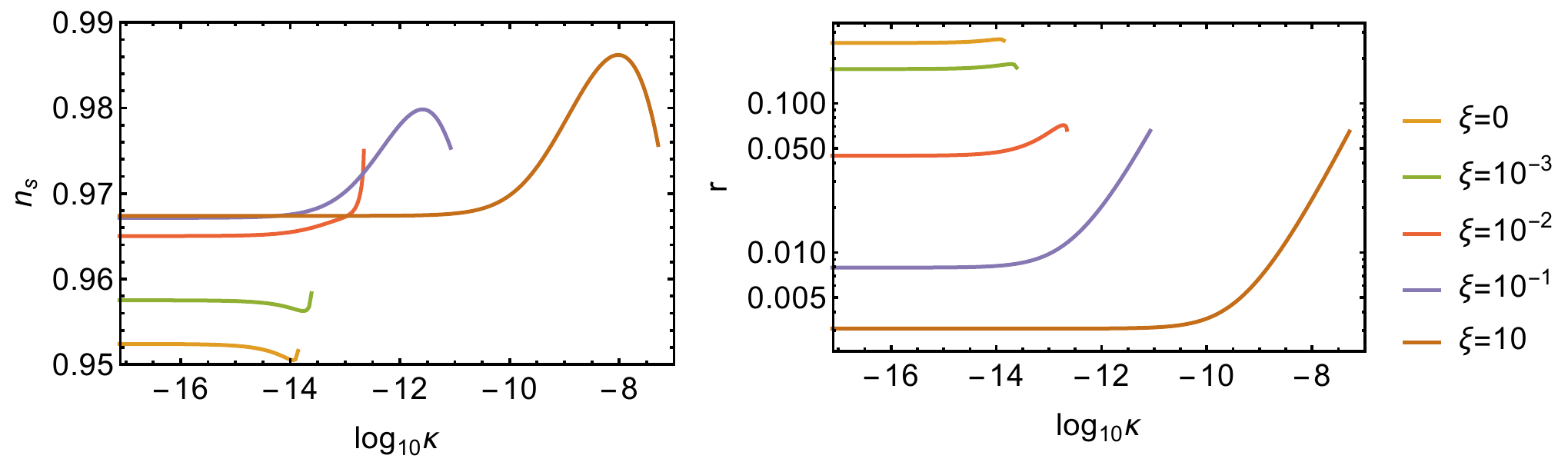}
	\end{center}\vspace*{-.5cm}
	\caption{For prescription II and inflaton coupling to bosons, the
change in $n_s$ and $r$ as a function of $\kappa$ is plotted for selected
$\xi$ values. The pink (red) contour in the top figure corresponds to the
95\% (68\%) CL contour based on data taken by the Keck Array/BICEP2 and
Planck collaborations \cite{Ade:2018gkx}. The red points
show the maximum $\kappa$ values.
These values, increasing with $\xi$, are also written in the figure.}
	\label{7.12}
\end{figure}

For prescription II and inflaton coupling to bosons, \fig{7.3} shows the
region in the $\xi$ and $\kappa$ plane where $n_s$ and $r$ values are
compatible with the current data. \Fig{7.12} shows how $n_s$ and $r$ values
change with $\kappa$ for chosen $\xi$ values. The $\kappa_{\mathrm{max}}$
values, that is, the maximum $\kappa$ values that allow a simultaneous
solution of eqs. (\ref{perturb2}), (\ref{efold2}) and (\ref{highN}) are
also shown. From the figures we see that
for $\xi\gtrsim10^{-2}$, the $n_s$ and $r$ values approach the linear
potential predictions $n_s\approx1-3/(2N)$ and $r\approx4/N$ as $\kappa$
approaches $\kappa_{\mathrm{max}}$.  This result is not surprising since
for large enough $\xi$ values $\xi\phi^2\gg1$ during inflation, in which
case the Einstein frame potential written in terms of the canonical scalar
field using eqs. (\ref{Zphi}) and (\ref{redefine}) contains a linear term
which eventually dominates as the value of $\kappa$ is increased. This
approach to the linear potential predictions was also noted in refs.
\cite{Martin:2013tda,Rinaldi:2015yoa,Racioppi:2018zoy}. Similarly to the
prescription I case for inflaton coupling to fermions, significant change in the $n_s$
and $r$ values only occur when $\kappa$ becomes the same order of magnitude
as $\kappa_{\mathrm{max}}$.

\begin{figure}[t]
	\begin{center}
		\includegraphics[width=7.5cm]{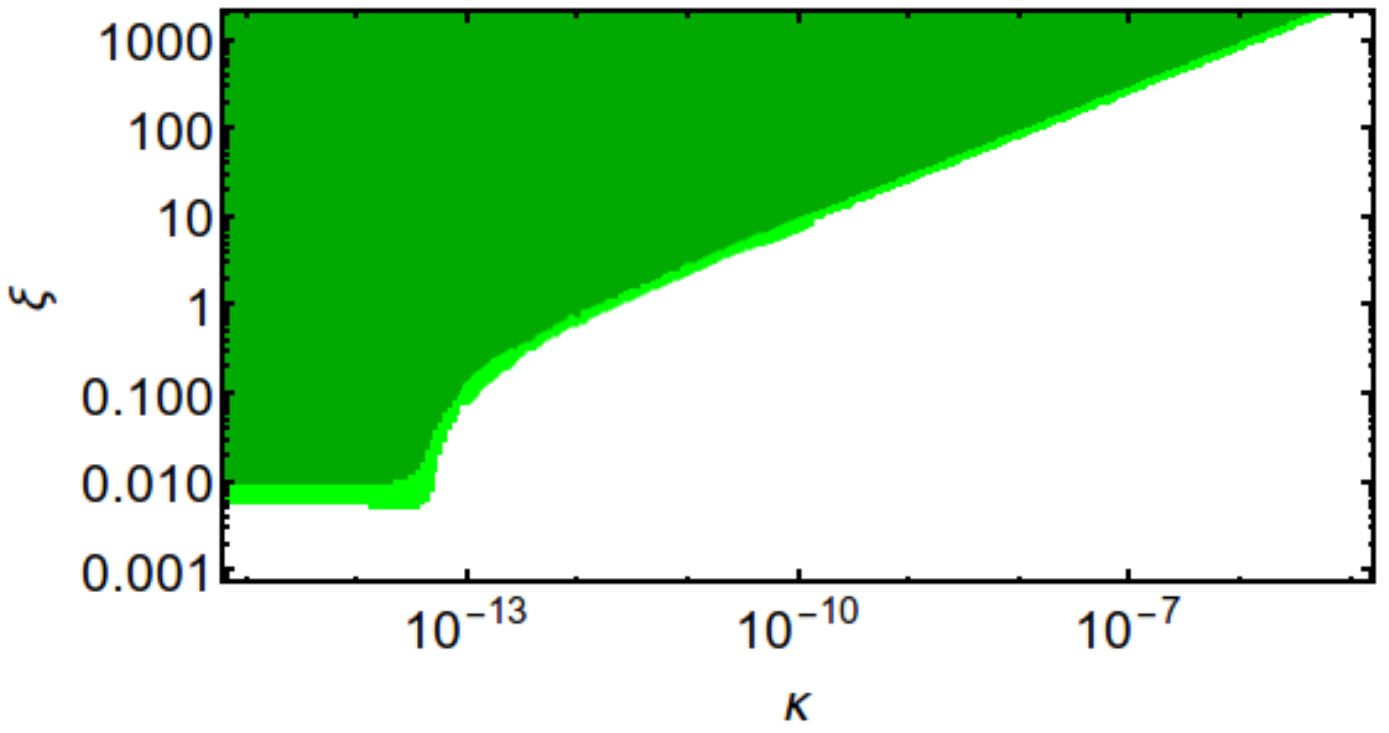}\\
		\includegraphics[width=14cm]{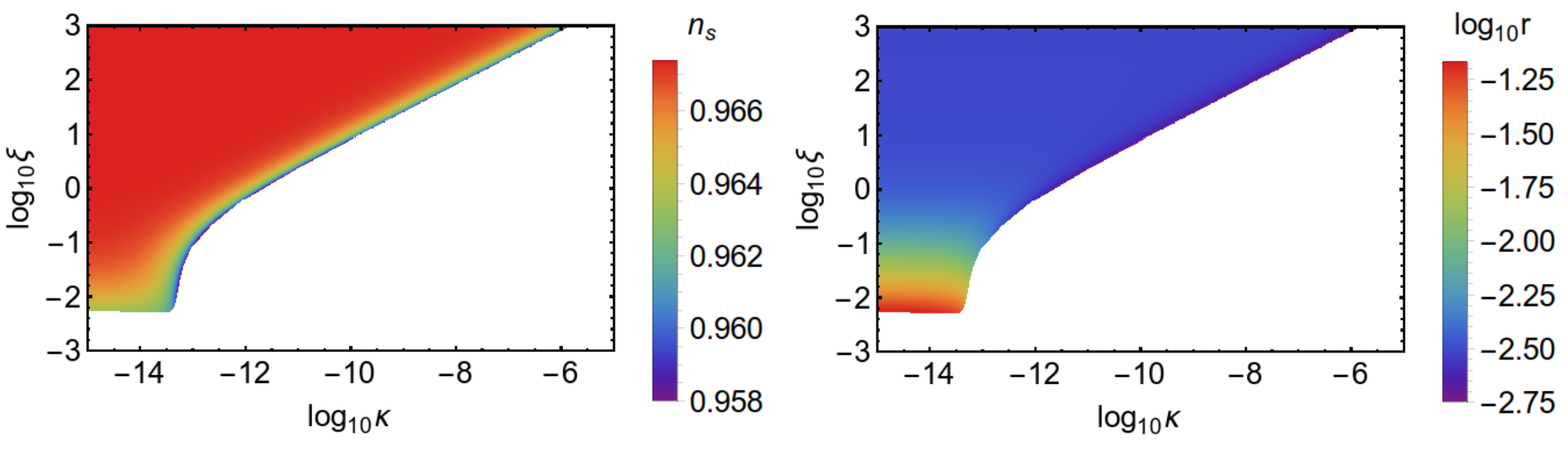}
	\end{center}\vspace*{-.5cm}
	\caption{For prescription II, inflaton coupling to fermions and
first branch solutions, the top figure shows in light green (green) the
regions in the $\xi$--$\kappa$ plane for which $n_s$ and $r$ values are
within the $95\%$ $(68\%)$ CL contours based on data taken by the Keck
Array/BICEP2 and Planck collaborations \cite{Ade:2018gkx}.  Bottom figures
show $n_s$ and $r$ values in these regions.}
	\label{7.1}
\end{figure}

\begin{figure}[b]
	\begin{center}
		\includegraphics[width=10cm]{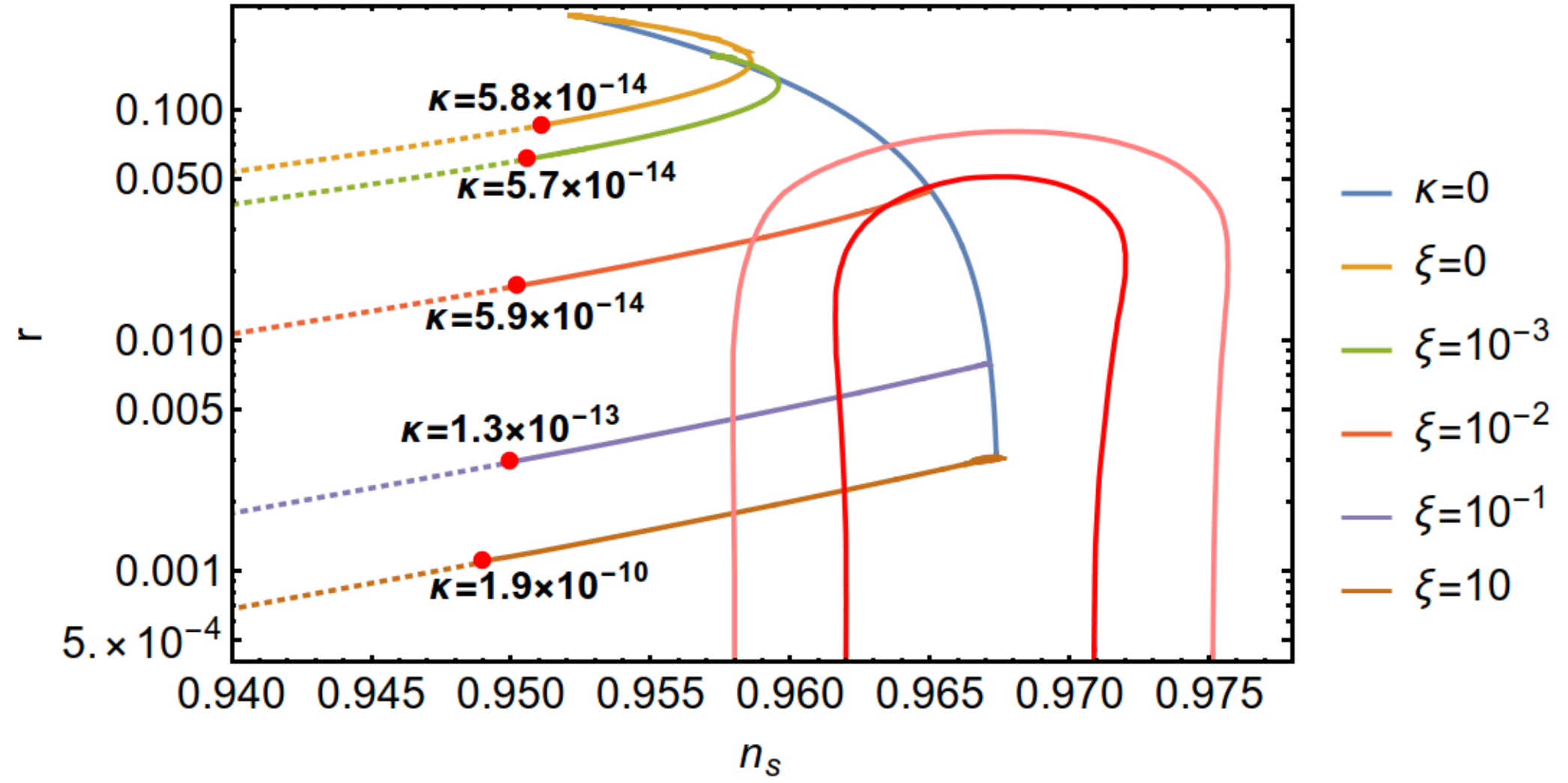}
		\includegraphics[width=14cm]{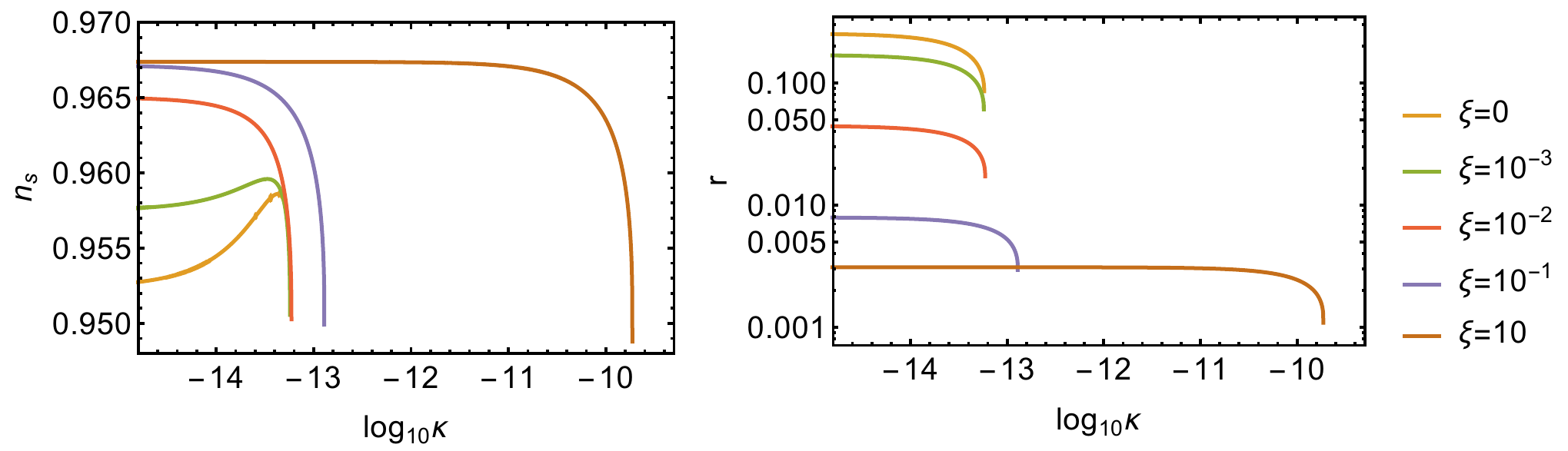}
	\end{center}\vspace*{-.5cm}
	\caption{For prescription II and inflaton coupling to fermions, the
change in $n_s$ and $r$ as a function of $\kappa$ is plotted for selected
$\xi$ values. The pink (red) contour in the top figure corresponds to the
95\% (68\%) CL contour based on data taken by the Keck Array/BICEP2 and
Planck collaborations \cite{Ade:2018gkx}. The solid (dotted) portions of
the curves correspond to first (second) branch of solutions. The red points
show the maximum $\kappa$ values where the two branch of solutions meet.
These values are also written in the figure. The
bottom figures only show the first branch solutions.}
	\label{7.14}
\end{figure}

Similarly to the prescription I case, there are also two branch of solutions for
prescription II and inflaton coupling to fermions. \Fig{7.1} shows the
region in the $\xi$ and $\kappa$ plane where $n_s$ and $r$ values are
compatible with the current data, for the first branch of solutions. The
second branch of solutions are not compatible with the current data at any
value of $\xi$ or $\kappa$. \Fig{7.14} shows how $n_s$ and $r$ values
change with $\kappa$ and the $\kappa_{\mathrm{max}}$ values for chosen
$\xi$ values. Again,  the first branch solutions move from the red points
towards the $\kappa=0$ curve as $\kappa$ decreases, whereas  the second
branch solutions move towards small $n_s$ values.

\enlargethispage{\baselineskip}

Finally we note that our results for prescription II and inflaton coupling
to fermions overlap and agree with those of ref. \cite{Okada:2010jf}.

\clearpage

\clearpage
%%%%%%%%%%%%%%%%%%%%%%%%%%%%%%%%%%%%%%%%%%%%%%%%%%%%%%%%%%%%%%%%

\section{Conclusion} \label{sec:conc}

In this paper we revisited the non-minimal quartic inflation model
consisting of a quartic potential and a coupling term $\xi \phi^2
R$ between the Ricci scalar and the inflaton, first
reviewing the tree level case without any radiative corrections in
\sektion{sec:quartic}. We noted that the approximate analytical expressions
in ref.  \cite{Bezrukov:2013fca} can be improved by using the
$W_{-1}(-x)\approx\ln x-\ln(-\ln x)$ approximation for the Lambert
function.

Two prescriptions used in the literature to take into account the
radiative corrections to the potential were briefly discussed in
\sektion{sec:radiative}. We then numerically investigated the effect of the
radiative corrections on the inflationary observables $n_s$ and $r$ due to
inflaton coupling to bosons or fermions in \sektion{sec:p1} for
prescription I and in \sektion{sec:p2} for prescription II.

Generally, we observed that while the radiative corrections prevent
inflation with a sufficient duration after a $\xi$ dependent maximum value $\kappa_{\mathrm{max}}$ of the coupling
parameter $\kappa$ defined by \eq{kappatanim}, they don't change $n_s$ and
$r$ values significantly unless $\kappa$ is the same order of magnitude as
$\kappa_{\mathrm{max}}$. For the prescription I and coupling to bosons
case, in contrast to the other cases, we found that eqs.
(\ref{perturb2}), (\ref{efold2}) and (\ref{highN}) can be simultaneously
satisfied for arbitrarily large values of $\kappa$. However, as explained
in \sektion{sec:p1}, we regard this result as an artifact of the
approximation we used for the potential.

The two prescriptions for the radiative corrections lead to significantly
different potentials in the limit $\xi\phi^2\gg1$, corresponding to
$\xi\gg1/(8N)$. For prescription I, the plateau type structure of the
potential remains intact in this limit. As a result, for the same $\kappa$
value the effect of radiative corrections is milder compared to the results
obtained using prescription II. This difference is also reflected in the
$\kappa_{\mathrm{max}}$ values. For example, if inflaton couples to
fermions and $\xi=10$, $\kappa_{\mathrm{max}}$ is $2.2\times10^{-8}$
($1.9\times10^{-10}$) using prescription I (prescription II). Such
differences suggest the neeed for further work on the theoretical
motivations of these prescriptions used in the literature to calculate the
observational parameters.

%%%%%%%%%%%%%%%%%%%%%%%%%%%%%%%%%%%%%%%%%%%%%%%%%%%%%%%%%%%%%%%%

\section*{Acknowledgements} This work is supported by T\"UB\.ITAK (The
Scientific and Technological Research Council of Turkey) project number
116F385.

%\clearpage

%to prevent pagebreaks in between references:
\makeatletter
\interlinepenalty=10000

%comment out the following lines before submitting to arXiv
%\bibliographystyle{JHEP.bst} 
%\bibliography{quartic_radiative.bib}

%uncomment the following line before submitting to arXiv
%(skip bibtex and directly use bbl file)
\bibliography{quartic_radiative_v3}

\makeatother

\end{document}